\documentstyle[aps,twocolumn]{revtex}

\begin{document}
\title{\begin{flushright}
\footnotesize{CECS-PHY-99/14}\\
\footnotesize{gr-qc/9909015}
\end{flushright} Conserved charges for gravity with locally AdS asymptotics}
\author{Rodrigo Aros$^{1,4}$, Mauricio Contreras$^{1,4}$, Rodrigo Olea$^{1,3}$,
Ricardo Troncoso$^{1,2}$ and Jorge Zanelli$^{1,2}$}
\address{$^{1}$Centro de Estudios Cient\'{\i }ficos de Santiago, Casilla
16443,Santiago, Chile.\\
$^{2}$Departamento de F\'{\i }sica, Universidad de Santiago de Chile,
Casilla 307, Santiago 2, Chile.\\
$^{3}$Departamento de F\'{\i }sica, FCFM, Universidad de Chile, Casilla
487-3, Santiago, Chile. \\
$^{4}$Universidad Nacional Andr\'{e}s Bello, Sazie 2320, Santiago, Chile.}
\maketitle

\begin{abstract}
{\bf Abstract. }A\ new formula for the conserved charges in 3+1 gravity for
spacetimes with local AdS asymptotic geometry is proposed. It is shown that
requiring the action to have an extremum for this class of asymptotia sets
the boundary term that must be added to the Lagrangian as the Euler density
with a fixed weight factor. The resulting action gives rise to the mass and
angular momentum as Noether charges associated to the asymptotic Killing
vectors without requiring specification of a reference background in order
to have a convergent expression. A consequence of this definition is that
any negative constant curvature spacetime has vanishing Noether charges.
These results remain valid in the limit of vanishing cosmological constant.
\end{abstract}

\medskip 

Noether theorem is the time honored way of constructing conserved charges
for any physical system provided its symmetries have been identified.
Nevertheless, it is puzzling that the straightforward application of
Noether's theorem to gravitation theory does not seem to lead to the
expected expressions for mass and angular momentum as the conserved charges
associated to time translations an spatial rotations. Applying Noether's
theorem for diffeomorphisms to the Einstein-Hilbert action in the second
order formalism yields an expression for the conserved charges as a surface
integral first proposed by Komar \cite{Komar}, (see, e.g.\cite{Wald}). This
expression, however, requires different normalization factors for the mass
and the angular momentum in order to reproduce the weak field approximation.
The generalization of Komar's formula when a negative cosmological constant
is introduced, requires the further substraction of a background
configuration in order to render a finite result.

The search for a general formula which allows to compute the mass (and other
conserved charges) in General Relativity as a surface integral has proceeded
through a series of improvements.

In the Hamiltonian framework, Arnowitt, Deser and Misner found a conserved
charge which yields the right normalization \cite{ADM}. As a follow up in
this scheme, Abott and Deser generalized this result in case of negative
cosmological constant \cite{Abott-Deser}.

An expression which extends and formally justifies this treatment was given
by Regge and Teitelboim who showed that the conserved charge is the boundary
term that must be added to the Hamiltonian constraint so that its variation
be well-defined \cite{Regge-Teitelboim}. Hawking and Horowitz observed that
this boundary term in the Hamiltonian can be obtained from an action
principle appropriate for Dirichlet boundary conditions for the metric \cite
{Hawking-Horowitz}. This action, however, depends explicitly on the
background solution.

A common feature of these approaches is the need of choosing a reference
background with suitable matching conditions. These conditions depend on the
asymptotic and topological properties corresponding to the class of metrics
involved. However, such a choice is not necessarily unique, and in some
cases it might even be impossible to find one.

As it occurs in Yang-Mills theories, there are cases of physical interest
where one knows the asymptotic behavior of the curvature, but has no
information about the metric at infinity. In those instances, there is no
unique background substraction that provides a general expression for the
conserved charges which works properly for all solutions with the same
asymptotic curvature. In particular, background-dependent methods are
inappropriate to deal with every kind of asymptotically locally AdS ({\bf %
ALAdS}) solutions.

One way to deal with this problem was proposed by Ashtekar and Magnon \cite
{Ashtekar-Magnon}, who use conformal techniques to construct an expression
for the `conserved' quantities in asymptotically AdS spaces. The results
correspond to those obtained by Hamiltonian methods \cite
{Henneaux-Teitelboim}. In this construction there is no mention of an
action, thus in principle, it is not straightforward how to relate these
conserved quantities and those associated with the symmetries of an action
in the Lagrangian formalism.

Another way, inspired by the AdS/CFT conjecture \cite{Maldacena} has been
recently proposed \cite{BK}. This proposal considers the Einstein-Hilbert
action with a negative cosmological constant plus the Gibbons-Hawking
boundary term --needed so that the variational principle be well defined for
metrics satisfying Dirichlet boundary conditions--, plus an additional
counterterm action which depends on the induced metric at the boundary.
Thus, using the Brown-York quasilocal energy definition \cite{Brown-York}
the sought formula is obtained. However, there is no clear a priori insight
as to which counterterm must be added. Any functional of the boundary metric
can be used as counterterm action (see, e.g., \cite{EJM,Mann}). The only
natural criteria seem to be requiring covariance and convergence of the
expression. The result is a series which needs to be corrected as one
increases the complexity of the solution.

In what follows it is shown that using the first order formalism allows to
write down an action principle which has an extremum for any ALAdS solution.
A regularized, background-independent expression for the conserved charges
is obtained by a straightforward application of Noether's theorem.

Consider the Einstein-Hilbert action for gravity with negative cosmological
constant in 3+1 dimensions. In the first order formalism, where the vierbein
and the spin connection are varied independently, the action reads

\begin{equation}
I=\frac{\kappa }{2}\int_{M}\epsilon _{abcd}(R^{ab}e^{c}e^{d}+\frac{l^{-2}}{2}%
e^{a}e^{b}e^{c}e^{d})+B,  \label{action}
\end{equation}
where $\kappa =1/(16\pi G)$ and the cosmological constant is $\Lambda =-%
\frac{3}{l^{2}}$. Here $B$ denotes a boundary term, which should be a
functional of the fields at infinity. The Noether currents and charges are
sensitive to this boundary term, and $B$ can be fixed by requiring that $%
I_{G}$ have an extremum on geometries which are ALAdS. Varying the action
yields

\begin{equation}
\delta I=\kappa \int_{M}\epsilon _{abcd}[\bar{R}^{ab}e^{c}\delta
e^{d}-T^{a}e^{b}\delta \omega ^{cd}]+\int_{\partial M}\Theta  \label{deltaI}
\end{equation}
where the boundary term in (\ref{deltaI}) is

\begin{equation}
\int_{\partial M}\Theta =\frac{\kappa }{2}\int_{\partial M}\epsilon
_{abcd}e^{a}e^{b}\delta \omega ^{cd}+\delta B,  \label{Theta}
\end{equation}
$T^{a}=De^{a}$\ is the torsion two-form and $\bar{R}%
^{ab}:=R^{ab}+l^{-2}e^{a}e^{b}$. The first integral in (\ref{deltaI}) is
zero on shell and therefore, the action has an extremum, provided $\Theta $
vanishes on the classical solution. Assuming that the spacetime is ALAdS ($%
\bar{R}^{ab}=0$ at $\partial M$), this condition is satisfied by

\begin{equation}
\delta B=\frac{\kappa l^{2}}{2}\int_{\partial M}\epsilon _{abcd}R^{ab}\delta
\omega ^{cd}.  \label{deltaBT}
\end{equation}
The r.h.s. of (\ref{deltaBT}) can be easily recognized as $\frac{\kappa l^{2}%
}{4}$ times the variation of the Euler (Gauss-Bonnet) density,

\[
\delta \left[ \int_{M}\epsilon _{abcd}R^{ab}R^{cd}\right] =2\int_{M}d\left[
\epsilon _{abcd}R^{ab}\delta \omega ^{cd}\right] . 
\]
Thus, the action is fixed up to a constant as

\begin{equation}
I=\frac{\kappa l^{2}}{4}\int_{M}\epsilon _{abcd}\bar{R}^{ab}\bar{R}^{cd}{}.
\label{BI}
\end{equation}

Either by direct variation of (\ref{BI}) or by substitution of (\ref{deltaBT}%
) into (\ref{Theta}), the term $\Theta $ is given by 
\begin{equation}
\Theta (\omega ^{ab},\delta \omega ^{ab},e^{a})=\frac{\kappa l^{2}}{2}%
\epsilon _{abcd}\delta \omega ^{ab}\bar{R}^{ab},  \label{ThetaCondition}
\end{equation}
up to an exact form, which is annihilated by the requirement $\Theta =0$ on $%
\partial M$. On the other hand, this requirement, in turn, could be
satisfied either by a Dirichlet condition on the spin connection ($\delta
\omega ^{ab}=0$) or by a sort of Neumann condition which imposes the
boundary of the spacetime to have negative constant curvature ($\bar{R}%
^{ab}=0$). If the only boundary is the asymptotic region, this last
condition is equivalent to require that spacetime be ALAdS.

The action (\ref{BI}) is manifestly invariant under diffeomorphisms, and
therefore, Noether's theorem provides a conserved current associated with
this invariance, given by

\begin{equation}
\ast J=-\Theta (\omega ^{ab},e^{a},\delta \omega ^{ab})-I_{\xi }L,
\label{JNoether}
\end{equation}
where the Lagrangian $L$ and $\Theta $ are defined in (\ref{BI}) and in (\ref
{ThetaCondition}) respectively and $I_{\xi }$ stands for the contraction
operator that lowers the degree of a $p$-form by one \cite{Footnote}. For a
diffeomorphism, $\delta \omega ^{ab}=-{\cal L}_{\xi }\omega ^{ab}$, and by
virtue of (\ref{JNoether}), the conserved current can be locally written as 
\begin{equation}
\ast J=\frac{\kappa l^{2}}{2}d(\epsilon _{abcd}I_{\xi }\omega ^{ab}\bar{R}%
^{cd}),  \label{Current}
\end{equation}
then, the conserved charge $Q(\xi )=\int_{\Sigma }*J$ is expressed as a
surface integral. The final expression is

\begin{equation}
Q(\xi )=\frac{\kappa l^{2}}{2}\int_{\partial \Sigma }\epsilon _{abcd}I_{\xi
}\omega ^{ab}\bar{R}^{cd}  \label{QAdS}
\end{equation}
where $\Sigma $ a spatial section of the manifold.

In sum, the condition that the action should have an extremum for any ALAdS
spacetime requires the addition of the Euler density as the boundary term
with the fixed weight $-\frac{3\kappa }{4\Lambda }$. This prescription
provides a simple and useful tool to compute charges in General Relativity.
For instance, the energy is obtained when $\xi $ is a time-like Killing
vector.

Although the ALAdS condition implies $\bar{R}^{cd}=0$ at the boundary, the
integrand of (\ref{QAdS}) approaches a finite value on $\partial \Sigma $.
In fact, standard boundary conditions \cite{Henneaux-Teitelboim} guarantee
the convergence of the charge for spatially localized configurations. Under
those same conditions, the contribution to the action from spatial infinity
can be directly shown to be finite as well. In the following examples the
convergence of (\ref{QAdS}) will be explicitly tested for geometries of
different topologies with timelike and spacelike Killing vectors.

\begin{itemize}
\item  {\bf Kerr-AdS}
\end{itemize}

The Kerr-AdS geometry in Boyer-Lindquist-type coordinates \cite{Carter} can
be expressed choosing the local Lorentz frame

\[
\begin{array}{ll}
e^{0}=\frac{\sqrt{\Delta _{r}}}{\Xi \rho }(dt-a\sin ^{2}\theta d\varphi ), & 
e^{1}=\rho \frac{dr}{\sqrt{\Delta _{r}}}
\end{array}
, 
\]

\begin{equation}
\begin{array}{ll}
e^{2}=\rho \frac{d\theta }{\sqrt{\Delta _{\theta }}}, & e^{3}=\frac{\sqrt{%
\Delta _{\theta }}}{\Xi \rho }\sin \theta (adt-(r^{2}+a^{2})d\varphi )
\end{array}
,  \label{KerrTetrad}
\end{equation}
with $\Delta _{r}=(r^{2}+a^{2})\left( 1+\frac{r^{2}}{l^{2}}\right) -2mr$, $%
\Delta _{\theta }=1-\frac{a^{2}}{l^{2}}\cos ^{2}\theta $, $\Xi =1-\frac{a^{2}%
}{l^{2}}$ and $\rho ^{2}=r^{2}+a^{2}\cos ^{2}\theta $. The mass and the
angular momentum are found evaluating the charge (\ref{QAdS}) for the
Killing vectors $\frac{\partial }{\partial t}$ and $\frac{\partial }{%
\partial \varphi }$, respectively 
\begin{equation}
\begin{array}{ll}
Q(\frac{\partial }{\partial t})=\frac{m}{\Xi ^{2}}; & Q(\frac{\partial }{%
\partial \varphi })=\frac{ma}{\Xi ^{2}}
\end{array}
,  \label{MJKerr}
\end{equation}
in agreement with Ref. \cite{Henneaux-Teitelboim}.

Note that formula (\ref{QAdS}) yields the right values for the mass and
angular momentum without need for adjusting the normalization factors as it
happens using the Komar potential.

\begin{itemize}
\item  {\bf (Un)wrapped black string}
\end{itemize}

Another kind of ALAdS spacetime, but with different topology and different
asymptotic behavior is given by the following black string geometry (see,
e.g. \cite{Lemos,Horowitz-Myers})

\begin{equation}
ds^{2}=-(\frac{r^{2}}{l^{2}}-\frac{b}{r})dt^{2}+\frac{dr^{2}}{(\frac{r^{2}}{%
l^{2}}-\frac{b}{r})}+r^{2}(dy_{1}^{2}+dy_{2}^{2}).  \label{Horowitz}
\end{equation}
In this metric, at least one of the transverse directions $y_{i\text{ }}$%
should be compactified, so that the parameter $b$ cannot be changed by
rescaling the coordinates. Thus, the only non-vanishing conserved charge is
the mass, as is depicted below

\begin{equation}
\begin{array}{ll}
Q(\frac{\partial }{\partial t})=\frac{bV}{8\pi }\;, & Q(\frac{\partial }{%
\partial y_{1}})=Q(\frac{\partial }{\partial y_{2}})=0,
\end{array}
\label{MJLemos}
\end{equation}
where $V$ stands for the two-dimensional transverse volume. Then, the
constant $b=\frac{8\pi M}{V}$ is a density of mass, unless both transverse
directions be compact. These results are in agreement with the computation
of Horowitz and Myers, using a background-dependent method \cite
{Horowitz-Myers}.

\begin{itemize}
\item  {\bf Taub-NUT \& Bolt}
\end{itemize}

Finally, we consider the Euclidean instanton solutions known as Taub-NUT and
Taub-Bolt (see, e.g. \cite{Hawking-Hunter-Page}), whose metrics can be
written as \cite{NUT-Bolt}

\begin{eqnarray}
ds^{2} &=&\frac{E}{4}\left[ \frac{F(r)}{E(\frac{r^{2}}{l^{2}}-1)}(d\tau +lE^{%
\frac{1}{2}}\cos \theta d\phi )^{2}+\right.  \nonumber \\
&&\left. \frac{4(\frac{r^{2}}{l^{2}}-1)}{F(r)}dr^{2}+(r^{2}-l^{2})(d\theta
^{2}+\sin ^{2}\theta d\phi ^{2})\right] .  \label{TaubNut-Bolt}
\end{eqnarray}
Using formula (\ref{QAdS}), the following results for the mass and angular
momentum are obtained for each case:

{\bf NUT:} 
\begin{equation}
\begin{array}{ll}
Q_{N}(\frac{\partial }{\partial \tau })=\frac{l}{2}E^{\frac{1}{2}}(1-E)\;, & 
Q_{N}(\frac{\partial }{\partial \phi })=0
\end{array}
\label{MJNut}
\end{equation}

{\bf Bolt:}

\begin{eqnarray}
&& 
\begin{array}{l}
Q_{B}(\frac{\partial }{\partial \tau })=\frac{l}{16s}E^{\frac{1}{2}%
}[Es^{4}+(4-6E)s^{2}-(3E-4)],
\end{array}
\nonumber \\
&& 
\begin{array}{l}
Q_{B}(\frac{\partial }{\partial \phi })=0
\end{array}
.  \label{MJBolt}
\end{eqnarray}
Using Taub-NUT as the background configuration, Hawking, Hunter and Page
computed the mass for the Taub-Bolt geometry. Substracting (\ref{MJNut})
from (\ref{MJBolt}), one obtains

\[
\Delta M=\frac{l}{16s}E^{\frac{1}{2}}[Es^{4}+(4\!-\!6E)s^{2}\!-\!(3E\!-%
\!4)+8(E\!-\!1)s], 
\]
which agrees with \cite{Hawking-Hunter-Page,refEJM}.

From the previous examples it is possible to observe that in all cases the
mass and other conserved charges can be computed from the formula (\ref{QAdS}%
), without normalizing the Killing vector as is often found in standard
approaches \cite{Killing}.

A remarkable feature of formula (\ref{QAdS}) is that the following general
statement can be derived from it:\newline

{\em For any negative constant curvature manifold, the Noether charges
associated with diffeomorphisms vanish identically.} {\em As a consequence,
any locally AdS spacetime which admits a set of global Killing vectors, has
all its associated conserved charges identically zero.}\newline

This means that any local AdS geometry possessing a time-like Killing vector
has vanishing mass. Considering that gravity has non-negative energy \cite
{Schoen-Yau-Witten}, the above statement implies that General Relativity has
a degenerate vacua: the only way to distinguish them would be through their
topological features.

An important problem in this context would be to identify all possible
vacua. It is easily seen that suitable vacua can be constructed as quotients
of the covering of $AdS$\ by a discrete group $\Gamma \;$that do not leave
timelike or spacelike fixed points, to avoid causal and conical
singularities, respectively. In fact, AdS spacetime is clearly not {\em the
unique} possible vacuum, since another choice is obtained from the metric (%
\ref{Horowitz}) with $b=0$, which corresponds to AdS spacetime with
identifications.

{\bf Discussion.} The Lagrangian in (\ref{BI}) can be understood as follows.
The most general $4$-form invariant under local Lorentz transformations,
constructed with the vielbein, the spin connection, and their exterior
derivatives, without using the Hodge dual ($*$) and even under parity is a
linear combination of the standard Einstein-Hilbert Lagrangian with a
cosmological term plus the Euler density. Hence, requiring that any ALAdS
spacetime be an extremum for the action fixes the weight factor of the Euler
density in terms of gravitational and cosmological constant as $-\frac{%
3\kappa }{4\Lambda }$. This is the key point in order to obtain the charge (%
\ref{QAdS}).

Since the Lagrangian is invariant under local Lorentz transformations,
substitution of $\delta \omega ^{ab}=-D\lambda ^{ab}$ in the general
expression for the Noether current (\ref{JNoether}) yields the conserved
charge

\begin{equation}
Q(\lambda ^{ab})=\frac{\kappa l^{2}}{2}\int_{\partial \Sigma }\epsilon
_{abcd}\lambda ^{ab}\bar{R}^{cd}.  \label{QLorentz}
\end{equation}
This expression is manifestly Lorentz covariant. On the other hand, in case
of diffeomorphism symmetry, $Q(\xi )$ given by (\ref{QAdS}), transforms
under local Lorentz rotations as

\begin{equation}
\delta _{\lambda }Q(\xi )=-\frac{\kappa l^{2}}{2}\int_{\partial \Sigma
}\epsilon _{abcd}{\cal L}_{\xi }\lambda ^{ab}\bar{R}^{cd}.  \label{DeltaQ}
\end{equation}
Thus, a sufficient condition to ensure the invariance of $Q(\xi )$ is, as
usual, the requirement ${\cal L}_{\xi }\lambda ^{ab}=0$ at the boundary.

Komar's expression for the charge and formula (\ref{QAdS}) differ
substantially although both are deduced from Noether's theorem for
diffeomorphisms. In the latter case, the charge comes from an action
containing an additional closed form and the entire procedure has been
conducted in first order. To make a more explicit comparison, it is
convenient to separate the charge (\ref{QAdS}) as 
\begin{equation}
Q(\xi )=K(\xi )+X(\xi )+E(\xi ),  \label{QSplit}
\end{equation}
where $K(\xi )=-\kappa \int_{\partial \Sigma }\nabla ^{\mu }\xi ^{\nu
}d\Sigma _{\mu \nu }$\ is Komar's integral, $X(\xi )=-\frac{\kappa }{2}%
\int\limits_{\partial \Sigma }\epsilon _{abcd}\Phi ^{ab}e^{c}e^{d}$\ --with $%
\Phi ^{ab}=e^{a\mu }{\cal L}_{\xi }e_{\mu }^{b}$-- is a contribution due to
the local Lorentz invariance, which does not exist in the second order
formalism \cite{X}, and finally, $E(\xi )=\frac{\kappa l^{2}}{2}%
\int\limits_{\partial \Sigma }\epsilon _{abcd}I_{\xi }\omega ^{ab}R^{cd}$,
which results from the Euler density added to the Einstein-Hilbert action.
This last term serves two purposes: it eliminates the divergent terms in the
explicit evaluation for the solutions, and corrects the normalization
factors. From this point of view, this new term regularizes the Noether
charges in ALAdS spacetimes. This can be illustrated in the
Schwarzschild-AdS solution. In this case $\Phi ^{ab}=0$\ and $\xi =\partial
_{t}$, so that the $X(\xi )$\ term vanishes and 
\begin{equation}
\begin{array}{ll}
K(\xi )=\frac{M}{2}+\lim\limits_{r\rightarrow \infty }\frac{r^{3}}{2l^{2}},
& E(\xi )=\frac{M}{2}-\lim\limits_{r\rightarrow \infty }\frac{r^{3}}{2l^{2}}.
\end{array}
\label{K}
\end{equation}
Hence, $Q(\xi )=M$, which is clearly unaffected whether the limit $%
l\rightarrow \infty $\ is performed or not, which means that the formula is
valid for any value of the cosmological constant, including $\Lambda =0$.
However the construction leading to (\ref{QSplit}) could not have carried
out if $\Lambda =0$\ from the start. Thus, the presence of a cosmological
constant can be thought of as a regulator for General Relativity with $%
\Lambda =0$.

An interesting issue is finding the generalization of the charge formula (%
\ref{QAdS}) for higher dimensional gravity theories which admit a well
defined variational principle for ALAdS solutions. Indeed, there is a
natural way to extend this result to any even dimension d%
\mbox{$>$}%
4 \cite{ACOTZd}. For odd dimensions, however, this extension is not
straightforward.

{\bf Acknowledgments}

We are grateful to J. Cris\'{o}stomo, J. Gamboa, A. Gomberoff, C. Mart\'{\i
}nez, F. M\'{e}ndez, M. Plyushchay, J. Saavedra and C. Teitelboim for many
enlightening discussions. We are specially thankful to M. Henneaux for many
helpful comments. This work was supported in part through grants 1990189,
1970151, 1980788, 2990018, 3960007, 3960009 and 3990009 from FONDECYT, and
grant 27-953/ZI-DICYT (USACH). The institutional support of Fuerza A\'{e}rea
de Chile, I.\ Municipalidad de Las Condes, and a group of Chilean companies
(AFP Provida, Business Design Associates, CGE, CODELCO, COPEC, Empresas
CMPC, GENER\ S.A., Minera Collahuasi, Minera Escondida, NOVAGAS and
XEROX-Chile) is also recognized. R.T. and J.Z. wish to express their
gratitude to Marc Henneaux, for kind hospitality in Brussels. Also R.A.,
R.O. and J.Z. wish to thank to the organizers of the ICTP Conference on
Black Hole Physics for hospitality in Trieste this summer.


\begin{references}
\bibitem{Footnote}  The contraction operator $I_{\xi }$ acts on a $p$-form

$\alpha _{p}=\frac{1}{p!}\alpha _{\mu ^{1}\cdot \cdot \cdot \mu ^{p}}dx^{\mu
^{1}}\cdot \cdot \cdot dx^{\mu ^{p}}$ as

$\;\;\;\;\;\;\;\;\;\;\;I_{\xi }\alpha _{p}:=\frac{1}{(p-1)!}\xi ^{\nu
}\alpha _{\nu \mu ^{1}\cdot \cdot \cdot \mu ^{p-1}}dx^{\mu ^{1}}\cdot \cdot
\cdot dx^{\mu ^{p-1}}$.

The Lie derivative can be written in terms of this operator as ${\cal L}%
_{\xi }=dI_{\xi }+I_{\xi }d$. Therefore, the following useful property holds
when it acts over the spin connection, ${\cal L}_{\xi }\omega ^{ab}=D(I_{\xi
}\omega ^{ab})+I_{\xi }R^{ab}$.
\bibitem{NUT-Bolt}  $F_{N}=E\frac{r^{4}}{l^{4}}+(4-6E)\frac{r^{2}}{l^{2}}%
+(8E-8)\frac{r}{l}+4-3E$ for the Taub-NUT case and

$F_{B}=E\frac{r^{4}}{l^{4}}+(4-6E)\frac{r^{2}}{l^{2}}+\left[ -Es^{3}+(6E-4)s+%
\frac{3E-4}{s}\right] \frac{r}{l}+4-3E$ for Taub-Bolt, for which $%
E=(2ks-4)/(3s^{2}-3)$.
\bibitem{refEJM}  The same expressions for the NUT and Bolt masses have been
found in Ref. \cite{EJM}.
\bibitem{Killing}  Once normalized, $\xi $ is no longer a Killing vector.
\bibitem{X}  Although it is always possible to choose a frame $(e^{a})$\
such that $\Phi ^{ab}=0$\ in an open neighborhood, the first order formalism
could provide interesting cases where there is a global obstruction which
makes $X(\xi )$\ non-trivial.
\end{references}
\end{document}